\documentclass[11pt]{article}
\usepackage[pdftex]{graphicx}
\usepackage{amsmath}    
\usepackage{amssymb}

\begin{document}

\begin{titlepage}
\begin{flushright}COLO-HEP-566 \end{flushright}
\vskip 1in

\begin{center}
{\Large{
The G\"odel-Schr\"odinger Spacetime and Stringy Chronology Protection
}}
\vskip 0.5in Charles Max Brown$^{a,b}$ and Oliver DeWolfe$^{b}$
\vskip 0.4in {\it $^{a}$ Department of Mathematics and Physics\footnote{Permanent address.} \\
Kentucky State University\\ Carver Hall 131, 400 E. Main St.\\
Frankfort, KY 40601, USA}
\vskip 0.2in {\it $^{b}$ Department of Physics \\ University of Colorado\\ 390 UCB \\ Boulder, CO 80309, USA}
\end{center}
\vskip 0.5in

\begin{abstract}

\noindent
We show that the null Melvin map applied to a rotational isometry of global anti-de Sitter space produces a spacetime with properties analogous to both the G\"odel and Schr\"odinger geometries.  The isometry group of this G\"odel-Schr\"odinger spacetime is appropriate to provide a holographic dual for the same non-relativistic conformal theory as the ordinary Schr\"odinger geometry, but defined on a sphere.  This spacetime also possesses closed timelike curves outside a certain critical radius.  We show that a holographic preferred screen for an observer at the origin sits at an interior radius, suggesting that a holographic dual description would only require the chronologically consistent region. Additionally, giant graviton probe branes experience repulson-type instabilities in the acausal region, suggesting a condensation of branes will modify the pathological part of the geometry to remove the closed timelike curves.

\end{abstract}

\end{titlepage}

\section{Introduction}

Geometries in general relativity can possess various  pathological features, and it is of interest how string theory, as a quantum mechanical theory of gravity, might or might not remove these pathologies.  Curvature singularities are the most well-known problematic features, but another interesting class is closed timelike curves (CTCs), which apparently enable time travel and render the usual notions of causal time evolution untenable.    Hawking conjectured  that no geometry initially free of closed timelike curves would develop them, the so-called Chronology Protection Conjecture \cite{Hawking:1991nk}; but whether a quantum theory of gravity might remove closed timelike curves that are already classically present is another question.

In a separate development, there has been considerable interest in using the AdS/CFT correspondence to build models of condensed matter physics (for reviews, see \cite{Hartnoll:2009sz, Herzog:2009xv, McGreevy:2009xe, Ammon:2010zz}).  The Schr\"odinger spacetime, formulated in \cite{Son:2008ye, Balasubramanian:2008dm} and embedded in string theory in \cite{Herzog:2008wg, Maldacena:2008wh, Adams:2008wt}, is of interest due to its isometry group matching the non-relativistic conformal symmetry \cite{Nishida:2007pj} of certain ultracold gases at unitarity, although a number of issues with the application of the geometry remain (for a recent discussion of some aspects, see for example \cite{Balasubramanian:2010uw, Guica:2010sw, Adams:2011kb}).   The Schr\"odinger spacetime may be generated from the Poincar\'e patch of anti-de Sitter space (AdS) through the so-called ``null Melvin twist" \cite{Alishahiha:2003ru, Gimon:2003xk}, generalized to a wide class of cases in \cite{Adams:2009dm} (see also \cite{Imeroni:2009cs}).  Nonzero temperature is implemented through acting the twist on a geometry containing a black hole.

The appearance of rotation in such spacetimes is interesting and potentially relevant towards a description of rotating ultracold gases; for a relativistic example, see \cite{Sonner:2009fk}.  
It is also interesting from the point of view of gravity to understand what results from acting the null Melvin twist on a broader class of spacetime metrics.
A natural step is then to apply the null Melvin map to a rotating black hole.  These black holes, however, live in global anti-de Sitter space, and the only spacelike isometries available to perform the Melvin map are rotational isometries, distinct from the translational symmetries used to generate the original Schr\"odinger spacetimes and its descendants.  

To understand the kind of spacetime that results, we study here the simplest case containing the essential features: the geometry resulting from acting the null Melvin twist on  global anti-de Sitter space using a rotational isometry.  Since global AdS holographically realizes the same field theory as the Poincar\'e patch of AdS, but living on a sphere instead of flat space, it is natural to hypothesize that the geometry that results in this case will realize the same non-relativistic field theory as the original Schr\"odinger spacetime, but living on a sphere.  We find the resulting geometry is a homogeneous spacetime with a large isometry group,  very similar to the symmetry group for the Schr\"odinger spacetime, with the differences in the algebraic structure suitable for the space to indeed be dual to the non-relativistic system living on a sphere.  Moreover, the original  Schr\"odinger spacetime can be recovered through a coordinate limit of blowing up the size of the sphere on the novel geometry.

However, there is a surprising feature of this new geometry: closed timelike curves exist.  These
CTCs pass through every point in the spacetime, but have a minimal radius.  This aspect is strongly reminiscent of G\"odel spacetimes, both the original geometry \cite{Godel:1949ga}
and its supersymmetric descendants (see for example \cite{Gauntlett:2002nw, Harmark:2003ud}).  Consequently the spacetime we generate shares properties of both Schr\"odinger and G\"odel systems,  and we refer to it as the G\"odel-Schr\"odinger spacetime.  One is thus led to ask: is this G\"odel-Schr\"odinger spacetime pathological due to the closed timelike curves, or does string theory mitigate them in some way?

We consider two distinct but complementary arguments suggesting that string theory does indeed avoid the difficulties of the closed timelike curves.
 First, one may consider a holographic screen in the spirit of the covariant holographic conjecture \cite{Bousso:1999xy,Bousso:1999cb}.  For ordinary global AdS space, this screen lives at the boundary, and hence it is natural to identify the boundary with the ``location" of the dual field theory degrees of freedom;  the entire spacetime is then holographically described by these modes.
 In the G\"odel-Schr\"odinger case, however, the screen sits at a finite radius, interior to the smallest closed timelike curves; a similar situation holds for the original G\"odel geometry and some of its supersymmetric descendants \cite{Boyda:2002ba}.  Thus the degrees of freedom can be taken to live at the screen, and only the interior (CTC-free) part of the geometry is associated  to the holographic dual description.

Moreover, it is possible to place a brane probe inside the chronologically challenged region; since a brane is a nonlocal object, it can sample an entire closed timelike curve and hence should be aware of the pathology.  It was described by Dyson \cite{Dyson:2003zn} how for certain black hole spacetimes with closed timelike curves,  probe branes may develop a repulson-type instability \cite{Behrndt:1995tr,Kallosh:1995yz,Cvetic:1995mx}
 which one would expect to be resolved by the branes' condensation, analogous to the enhan\c{c}on mechanism \cite{Johnson:1999qt}. 
A similar test was performed in the case of a supersymmetric G\"odel spacetime \cite{Drukker:2003sc}, where it was found that a BPS supertube probe develops a negative kinetic term in the CTC region, implying an analogous instability.
 In close analogy, we find that a D3-brane ``dual giant graviton" probe develops negative kinetic terms precisely in the acausal region, and thus would be expected to condense.  
 
 We thus hypothesize that while the geometry we obtain is a solution to type IIB supergravity, it is not a true solution to type IIB string theory. Incorporating the condensed brane backreaction, a new solution sourced by the condensed branes should exist at large radius, while the small-radius geometry would persist; according to the holographic screen argument, only this region would be necessary to describe the dynamics experienced by a given observer.  (For further work on chronology protection in string theory, see \cite{Herdeiro:2000ap}-\cite{Raeymaekers:2010re}.)  The null Melvin map acting on rotating black holes would then produce holographic duals of rotating non-relativistic conformal field theories, whose analogous CTCs are presumably removed by a similar mechanism.

In section~\ref{MelvinSec}, we review the null Melvin twist and the original Schr\"odinger spacetime.   In section~\ref{GlobalSec} we apply the null Melvin twist to global anti-de Sitter space to generate the G\"odel-Schr\"odinger spacetime, calculate the isometries and their algebra, and obtain ordinary Schr\"odinger as a limit.
We demonstrate the existence of closed timelike curves larger than a certain critical radius in section~\ref{ScreenSec},  compare the geometry to the G\"odel and supersymmetric G\"odel spacetimes, and demonstrate that like those geometries, an observer at the origin of G\"odel-Schr\"odinger is associated to a holographic screen living inside the critical radius for CTCs.  Finally, in section~\ref{BraneSec} we identify D3-brane giant graviton solutions in the G\"odel-Schr\"odinger geometry and show that they develop negative kinetic terms precisely at the critical radius, suggesting an instability towards condensation of branes.  In section~\ref{ConclusionSec} we conclude.

\section{The Null Melvin twist and the Schr\"odinger spacetime}
\label{MelvinSec}

\subsection{The twist}

The Null Melvin twist is an operation involving compositions of coordinate shifts and two T-dualities, which maps solutions to solutions in type IIB supergravity \cite{Alishahiha:2003ru, Gimon:2003xk}.   It was used by \cite{Herzog:2008wg, Maldacena:2008wh, Adams:2008wt} to generate the Schr\"odinger spacetime from ordinary anti-de Sitter space in the Poincar\'e patch, with an added black hole leading to the finite-temperature generalization.
The Melvin map was generalized to include its action on charged black holes and other, broader classes of spacetimes  in  \cite{Adams:2009dm, Imeroni:2009cs};
 our discussion follows \cite{Adams:2009dm}.  (For further developments generalizing Schr\"odinger spacetimes, see for example \cite{Theis:2010ws, Kim:2010tf, Banerjee:2011jb, Kraus:2011pf}.)  
 
 The null Melvin twist requires as input a ten-dimensional metric with timelike isometry $\partial_\tau$ and two spacelike isometries $\partial_y$ and $\partial_\chi$.  The solutions we are interested in have a product form\footnote{It is straightforward to replace $S^5$ with a more general compact Sasaki-Einstein space, but since our focus will be on phenomena in $M$, we will not do so.} $M \times S^5$, where $\tau$ and $y$ are timelike and spacelike coordinates on $M$, respectively, and $\chi$ is the Hopf coordinate on $S^5$:
\begin{eqnarray}
ds^2 = ds^2(M) + ds^2({\mathbb P}^2) + (d\chi + {\cal A})^2 \,,
\end{eqnarray}
where the 1-form ${\cal A}$ defines the fibration of $S^5$ over ${\mathbb P}^2$.
The initial solution also has nonzero $\tilde{F}_5$, 
\begin{eqnarray}
\tilde{F}_5 = {\cal F}_5 + * {\cal F}_5 \,, \quad \quad {\cal F}_5 = -4 {\rm vol}_M \,,
\end{eqnarray}
with other RR fields and the NSNS potential $B_2$ vanishing, and the dilaton constant $\Phi= \Phi_0$.

One may always write the five-dimensional metric as
\begin{eqnarray}
ds^2(M) = G_{\alpha \beta} e^\alpha e^\beta + G_{mn} dx^m dx^n \,,
\end{eqnarray}
with $\alpha, \beta = \tau, y$ and $m, n$ running over the remaining three coordinates, and $e^\alpha \equiv dx^\alpha + A^\alpha_m dx^m$; note the cross terms between the $\tau$ or $y$ and the other three coordinates are absorbed into $e^\alpha$, while a cross term between $\tau$ and $y$ is written as an explicit $G_{\tau y}$ component.

The null Melvin twist then produces a new string frame type IIB supergravity solution, characterized by a parameter $\beta$, given by the NSNS sector
\begin{eqnarray}
ds^2 &=& ds^2(M)' + ds^2({\mathbb P}^2) + e^{2V} (d\chi + {\cal A})^2 \,, \\
B_2 &=& A_M \wedge (d\chi + {\cal A}) \,, \quad \quad \Phi = \Phi_0 - {1 \over 2} \log K \,,
\end{eqnarray}
with the RR sector unchanged, where $K$ is a scalar function and $A_M$ is a 1-form on $M$:
\begin{eqnarray}
K &=& e^{-2V} = 1 + \beta^2 (G_{\tau\tau} + G_{yy} - 2 G_{\tau y}) \,, \\
 A_M &=& - \beta e^{2V} ( G_{y \alpha} - G_{\tau \alpha}) e^\alpha \,,
\end{eqnarray}
and the new five-dimensional metric is
\begin{eqnarray}
ds^2(M)' = {\beta^2 ||G_{\alpha\beta}|| \over K} (e^\tau + e^y)^2  + {1 \over K} G_{\alpha \beta} e^\alpha e^\beta + G_{mn} dx^m dx^n \,,
\end{eqnarray}
with $||G_{\alpha\beta}|| = G_{\tau\tau} G_{yy} - G_{\tau y}^2$.  In the limit $\beta \to 0$, the original spacetime is recovered.  One may also write the difference between the original metric and the new metric in terms of the one-form $A_M$:
\begin{eqnarray}
\label{MetricAM}
ds^2(M)' - ds^2(M) = - K A_M \otimes A_M \,.
\end{eqnarray}
The effective five-dimensional theory from dimensionally reducing on the (squashed) $S^5$  is a theory of a metric, a massive gauge field, and since $V = \Phi - \Phi_0$, a single scalar:
\begin{eqnarray}
2 \kappa_5^2 S_5 =  \int {\rm vol}_M\,  e^{-V} \left( R^{(5)} + 16 - 4 e^{2V} \right) - {1 \over 2} \int \left(e^{-3V} F_M \wedge *_5 F_M + 8 e^{-V}A_M \wedge *_5 A_M
\right)\,.
\end{eqnarray}
This action is a truncation of a three-scalar Lagrangian described in \cite{Maldacena:2008wh};
we have set the AdS scale to unity.  When $A_M = V = 0$, this is just 5D Einstein gravity with a cosmological constant.  The mode $V$ has no kinetic term in this string frame, but regains one in Einstein frame, with $G^E_{\mu\nu} \equiv e^{-2V/3} G_{\mu\nu}$:
\begin{eqnarray}
2 \kappa_5^2 S_{5,E} &=&  \int {\rm vol}_M\, \left( R^{(5)}_E + 16 e^{2V/3} - 4 e^{8V/3} \right)  \\ && - {1 \over 2} \int \left( {8 \over 3} dV \wedge *_E d V+
e^{-8V/3} F_M \wedge *_E F_M + 8 A_M \wedge *_E A_M
\right)\,. \nonumber
\end{eqnarray}

\subsection{The Schr\"odinger solution and its symmetries}

We next review the five-dimensional Schr\"odinger solution and its symmetry group.
Five-dimensional anti-de Sitter space in the Poincar\'e patch may be written as (again, the AdS scale is unity):
\begin{eqnarray}
\label{PAdS}
ds^2 = - r^2 d\tau^2 +  r^2 d\vec{X}^2+ {dr^2 \over r^2}  \,,
\end{eqnarray}
with $d\vec{X}^2 = dx^2 + dy^2 + dz^2$.  Defining
\begin{eqnarray}
e^{(\tau)} = d \tau \,, \quad \quad e^{(y)} = d y \,,
\end{eqnarray}
we may apply the null Melvin twist, resulting in the expressions
\begin{eqnarray}
K &=& 1 + \beta^2(- r^2  + r^2 )= 1 \,, \\
||G_{\alpha \beta} ||&=& G_{\tau\tau} G_{yy} = - r^4 \,,
\end{eqnarray}
and leading to the five-dimensional metric 
\begin{eqnarray}
ds^2 = - \beta^2 r^4 (d\tau + dy)^2 - r^2 d\tau^2 +  r^2 d\vec{X}^2+ {dr^2 \over r^2}  \,,
\end{eqnarray}
and the massive vector
\begin{eqnarray}
A_M = - \beta r^2 (dy + d\tau) \,.
\end{eqnarray}
Since $K=1$, the dilaton remains constant at $\Phi = \Phi_0$, so we need not distinguish string and Einstein frame metrics.
It is standard to reexpress this solution in terms of light cone coordinates,
\begin{eqnarray}
\label{LightCone}
t \equiv \beta(\tau + y)\,, \quad \quad \xi \equiv (y - \tau)/2\beta \,,
\end{eqnarray}
in terms of which we obtain the metric
\begin{eqnarray}
\label{PSchro}
ds^2 = - r^4 dt^2 + r^2 ( 2 dt d\xi + d\vec{x}^2) + {dr^2 \over r^2} \,,
\end{eqnarray}
with $d\vec{x}^2 = dx^2 + dz^2$, and the 1-form
\begin{eqnarray}
A_M = - r^2 dt \,.
\end{eqnarray}
This is the Schr\"odinger solution, so called because its isometries coincide with the non-relativistic Schr\"odinger conformal symmetry group with dynamical exponent $z=2$, in $d=2$ spatial dimensions.  The utility of non-relativistic conformal field theories (NRCFTs) to ultracold gases at unitarity was described in \cite{Nishida:2007pj}, and the dual geometry (\ref{PSchro}) was first described in \cite{Son:2008ye, Balasubramanian:2008dm}, and lifted to a solution of type IIB string theory in \cite{Herzog:2008wg, Maldacena:2008wh, Adams:2008wt}.
We will denote the Schr\"odinger solution as ${\cal S}_5$.

The Schr\"odinger spacetime ${\cal S}_5$ possesses nine isometries,
\begin{eqnarray}
\nonumber
H &=& \partial_t \,, \quad \quad N = - \partial_\xi \,, \quad \quad P_x = \partial_x \,, \quad \quad P_z = \partial_z \,, \quad \quad M_{xz} = z \partial_x - x \partial_z \,, \\
\label{SchroIsoms}
D &=& - 2 t\partial_t + r \partial_r -x \partial_x - z \partial_z \,, \quad \quad
K_x = x \partial_\xi - t \partial_x \,, \quad \quad K_z = z \partial_\xi - t \partial_z \,,  \\
C &=& t^2 \partial_t - {1 \over 2} \left( x^2 + z^2 + {1 \over r^2} \right) \partial_\xi - t r \partial_r + t x \partial_x + t z \partial_z \,,
\nonumber
\end{eqnarray}
corresponding in the dual field theory to the Hamiltonian $H$, particle number $N$, momentum generators $P_i$, $SO(2)$ rotations $M_{xz}$, Galilean boosts $K_i$, dilatations $D$ and special conformal transformations $C$.
Several subgroups of the algebra are noteworthy.  The commutators of $H$, $D$ and $C$ are,
\begin{eqnarray}
[D, H] = 2 i H \,, \quad \quad [C, H] = i D \,, \quad \quad [C, D] = 2 i C \,,
\end{eqnarray}
which upon defining
\begin{eqnarray}
J_{12} \equiv {1 \over 2} (H + C) \,, \quad \quad J_{01} \equiv {1 \over 2} (H-C) \,, \quad \quad J_{02} \equiv {D \over 2} \,,
\end{eqnarray}
form the generators $J_{ab}$ of an $SL(2,R) = SO(2,1)$ subalgebra
\begin{eqnarray}
 [J_{ab}, J_{cd} ] = i (\eta_{ac} J_{bd} - \eta_{bc} J_{ad} + \eta_{bd} J_{ac} - \eta_{ad} J_{bc}) \,,
\end{eqnarray}
with $a,b = 0,1,2$ and $\eta_{ab} \equiv {\rm diag}(-1,1,1)$.  Furthermore, in $d$ spatial dimensions the $M_{ij}$ constitute the generators of the standard $d$-dimensional rotational algebra $SO(d)$,
\begin{eqnarray}
 [M_{ij}, M_{kl} ] = i (\delta_{ik} M_{jl} - \delta_{jk} M_{il} + \delta_{jl} M_{ik} - \delta_{il} M_{jk}) \,,
\end{eqnarray}
with $d=2$ in our case, and the $SL(2,R)$ and $SO(d)$ subalgebras commute,
\begin{eqnarray}
[J_{ab}, M_{ij}] = 0 \,.
\end{eqnarray}
These factors are linked through the momentum $P_i$ and boosts $K_i$, which are manifestly each vectors under $SO(d)$ and may be combined into an object ${\cal P}_i \equiv (P_i, K_i)$ which is also a doublet under $SL(2,R)$:
\begin{eqnarray}
[J_{12}, {\cal P}_i] = -{i \sigma^1\over 2} {\cal P}_i \,, \quad \quad  [J_{01}, {\cal P}_i] = -{ \sigma^2\over 2} {\cal P}_i \,, \quad \quad
 [J_{02}, {\cal P}_i] = {i \sigma^3\over 2} {\cal P}_i \,.
\end{eqnarray}
Finally the $P_i$ and $K_i$ each commute with themselves, but the mutual commutator is centrally extended by $N$:
\begin{eqnarray}
[P_i, P_j] = 0 \,, \quad \quad [K_i, K_j] = 0 \,, \quad \quad  [K_i, P_j] &=& i \delta_{ij} N \,,
\end{eqnarray}
where $N$ commutes with everything.  Thus the Poincar\'e algebra may be thought of as $SL(2,R) \times SO(d) \times U(1)$ (the $U(1)$ factor being $N$) augmented by the spinor-vector ${\cal P}$ which commutes onto the $U(1)$.

\section{Melvinization of global AdS}
\label{GlobalSec}

We now turn to the focus of our study, the geometry generated by applying the null Melvin twist to a rotational isometry in global anti-de Sitter space, which for reasons that will become apparent we will call the five-dimensional G\"odel-Schr\"odinger geometry, or ${\cal GS}_5$.  Before describing this spacetime, we will review global AdS, its symmetries and holographic interpretation.

\subsection{Global anti-de Sitter space}

Anti-de Sitter space can be presented in a number of  coordinate systems, some of which span the whole manifold and some of which do not.  The Poincar\'e patch metric (\ref{PAdS}), while very useful for AdS/CFT applications, does not cover the entire manifold as geodesics exit it in finite proper time.  Global anti-de Sitter space, in contrast, is locally the same geometry, but presented in a coordinate system that covers the entire, geodesically complete spacetime.  In $d+1$ dimensions, the global AdS metric may be written
\begin{eqnarray}
ds^2 =  -(1 + r^2) d\tau^2 +  { dr^2 \over 1 + r^2} + r^2 d \Omega_{d-1}^2 \,,
\end{eqnarray}
where $d \Omega_{d-1}^2$ is the metric on the round $(d-1)$-sphere.

Anti-de Sitter space in $d+1$ dimensions possesses a $SO(d,2)$ isometry group, corresponding to the  (relativistic) conformal group of the dual $d$-dimensional field theory.  Different subgroups of $SO(d,2)$ are more apparent in  different coordinate presentations.  The Poincar\'e patch metric (\ref{PAdS}) makes manifest the foliation of the space with respect to ${\mathbb R}^{d-1,1}$ spanned by $\tau$ and the $\vec{X}$, and the corresponding $SO(d-1,1)$ Lorentz subgroup and the $d$ translations are clearly visible.  The boundary of the Poincar\'e patch is a copy of ${\mathbb R}^{d-1,1}$, and so the dual field theory lives on this space.  The remaining generators of $SO(d,2)$ correspond to the dilatation and special conformal transformations in the dual conformal field theory.  
The Schr\"odinger isometries of ${\cal S}_5$ constitute a nine-dimensional subgroup of $SO(4,2)$ preserving (among others) the isometries used in the Melvin map.

In global AdS, on the other hand, the $SO(d)$ subgroup of $SO(d,2)$ acts in the usual way on the $(d-1)$-sphere, while $\tau$-translations form the $SO(2)$ subgroup.\footnote{Constructing AdS as a hyperboloid in a $(d+2)$-dimensional space, $\tau$ emerges as a periodic coordinate, and one typically passes to the universal cover to generate an infinite temporal extent.  We will not be careful distinguishing the $SO(2)$ symmetry of compact time from the ${\mathbb R}$ of the universal cover.}  The space is a foliation by $S^{d-1} \times {\mathbb R}$ (the universal cover of $S^{d-1} \times S^1$), on which the boundary theory lives; thus $SO(d)$ and $\partial_\tau$ are the geometrical symmetries of the dual field theory, and the remaining generators of $SO(d,2)$ are the conformal enhancement.  

The AdS/CFT interpretation is that the Poincar\'e and global AdS cases are dual to the same CFT, only living on different spaces, ${\mathbb R}^{d-1,1}$ vs. $S^{d-1} \times {\mathbb R}$.  
Note that it is rather nontrivial that the geometric symmetries of the two cases (which are quite distinct for a non-conformal theory) are enhanced to the identical symmetry group $SO(d,2)$ by the additional conformal generators.
The natural question then arises: can one find something similar for the Schr\"odinger case, a kind of ``global Schr\"odinger" spacetime, which describes the same dual field theory but living on a sphere instead of a plane?\footnote{Another approach to ``global Schr\"odinger" is to take the Schr\"odinger solution and maximally continue its geodesics to obtain a complete manifold; this was considered in \cite{Blau:2009gd, Blau:2010fh}. It would be interesting to further illuminate the connection between the geometry studied there and the ${\cal GS}_5$ spacetime.}  As we shall see, the geometry resulting from applying the null Melvin twist to a rotational generator in global AdS seems an appropriate candidate for such a spacetime, though the full group is not exactly the same as in the ${\cal S}_5$ Schr\"odinger case.

\subsection{Null Melvin twist of global AdS}

For global $AdS_5$, we may take the coordinates to be 
\begin{eqnarray}
\label{GlobalAdS}
ds^2_{\rm Global} =  -(1 + r^2) d\tau^2 +  {dr^2 \over 1 + r^2} + {r^2 \over 4} \left( (d \psi + \cos \theta d\phi)^2 + d\theta^2 + \sin^2 \theta d \phi^2 \right) \,.
\end{eqnarray}
The coordinates for the $S^3$ emphasize the Hopf fibration of $S^1$, parameterized by $\psi$, over the $S^2$ parameterized by $\theta$ and $\phi$.

To apply the Melvin map, we must pick a pair of isometries.  Global AdS has all the same isometries as the Poincar\'e patch, so in principle one could perform the same Melvin twist that led to ${\cal S}_5$ (\ref{PSchro}) by using a ``translation" generator.  We are interested in the result if a rotation generator on the $S^3$ is used instead.  There are several motivations for exploring this possibility.  The first is finding a gravitational dual for the NRCFT on a sphere, as described above.  Another is understanding the implementation of rotation in the background of the NRCFT, for which we want to explore the action of the Melvin map on rotating black holes.  The most general rotating black hole in 5D AdS space preserves only three isometries, $\partial_\tau$, $\partial_\psi$ and $\partial_\phi$.  We may study the null Melvin twist applied to a rotational isometry of AdS as the simplest case of the class of spaces that include the Melvinization of the rotating black holes.

Even choosing to pick a rotational direction, there is a certain amount of freedom still in defining the isometry. Our philosophy will be to pick the isometry so as to make the resulting geometry resemble the Schr\"odinger solution as much as possible.
We will define the rotational isometry, the $y$ direction, as proportional to the Hopf fiber direction $\psi$:
\begin{eqnarray}
\label{ChoosePsi}
\psi= \alpha y \,,
\end{eqnarray}
where $\alpha$ is a constant, so that
\begin{eqnarray}
ds^2 =  -(1 + r^2) d\tau^2 + (1 + r^2)^{-1} dr^2 + {\alpha^2 r^2 \over 4} \left( (d y + {1 \over \alpha} \cos \theta d\phi)^2 + {1\over \alpha^2} d\theta^2 + {1 \over \alpha^2} \sin^2 \theta d \phi^2 \right) \,.
\end{eqnarray}
Then we have
\begin{eqnarray}
e^{(\tau)} = d\tau \,, \quad \quad e^{(y)} = d y + {1 \over \alpha} \cos \theta d\phi \,,
\end{eqnarray}
and thus 
\begin{eqnarray}
K &=& 1 + \beta^2 (G_{\tau \tau} + G_{yy} - 2 G_{\tau y}) = 1 + \beta^2 \left[- 1 +  \left( {\alpha^2 \over 4} - 1 \right) r^2\right]\,, \\
|| G_{\alpha\beta} || &=& G_{\tau \tau} G_{yy} - G_{\tau y}^2 = - {\alpha^2 \over 4} r^2 (1 + r^2) \,.
\end{eqnarray}
We see that to avoid $K$ going negative (and the string coupling becoming imaginary), we need $\alpha^2 \geq 4$.  To further constrain this parameter, we note that the Schr\"odinger solution ${\cal S}_5$ has a constant dilaton.  This may be achieved in the present case by demanding
\begin{eqnarray}
\alpha = 2 \,,
\end{eqnarray}
giving
\begin{eqnarray}
K &=& 1 - \beta^2 \,, \\
|| G_{\alpha\beta} || &=&  - r^2 (1 + r^2) \,.
\end{eqnarray}
Notice we still must require $\beta^2 < 1$ to avoid the unphysical result of vanishing or negative $K$.
As with ${\cal S}_5$, the resulting constant dilaton means we need not distinguish the Einstein and string frame metrics.
  Note that a more general choice of rotational isometry (\ref{ChoosePsi}) involving both $\psi$ and $\phi$ also results in a varying dilaton.

Then the Melvinized metric is
\begin{eqnarray}
ds^2 &=& - {\beta^2 \over 1 - \beta^2} r^2 (1+r^2) \left(d\tau +  {1 \over 2} ( d\psi + \cos \theta d\phi) \right)^2 \\
&& -{1 \over 1 - \beta^2} (1 + r^2) d\tau^2 + (1 + r^2)^{-1} dr^2 + {r^2 \over 4} \left( {1 \over 1 - \beta^2} (d \psi + \cos \theta d\phi)^2 + d\theta^2 + \sin^2 \theta d \phi^2 \right) \,, \nonumber
\end{eqnarray}
which we can also write as
\begin{eqnarray}
ds^2 &=&- (1 + r^2) {1 + \beta^2 r^2 \over 1- \beta^2} d\tau^2 - {\beta^2 r^2 (1+r^2)\over 1 - \beta^2} d\tau (d\psi + \cos \theta d\phi) + {dr^2 \over 1+r^2} \\
&& + {r^2 \over 4(1-\beta^2)} (1 -\beta^2 - \beta^2 r^2) (d\psi + \cos \theta d \phi)^2 + {r^2 \over 4}  d\theta^2 + {r^2 \over 4}  \sin^2 \theta d \phi^2 \,, \nonumber
\end{eqnarray}
along with the massive vector field and dilaton,
\begin{eqnarray}
A_M = - {\beta \over 1 - \beta^2} \left( (1 + r^2) d\tau + {r^2 \over 2} (d\psi + \cos \theta d \phi) \right) \,, \quad \quad
e^\Phi  = {e^{\Phi_0} \over \sqrt{1 - \beta^2}} \,.
\end{eqnarray}
We may also present the metric more compactly in terms of the massive vector field as in (\ref{MetricAM}),
\begin{eqnarray}
\label{MGAdS}
ds^2 = ds^2_{\rm Global} - {1 \over R_{\rm CTC}^2} \left( (1 + r^2) d\tau + {r^2 \over 2} (d\psi + \cos \theta d \phi) \right)^2 \,,
\end{eqnarray}
with $ds^2_{\rm Global}$ the global AdS metric given in (\ref{GlobalAdS}), and where we have defined
\begin{eqnarray}
\label{DefineRCTC}
R_{\rm CTC} \equiv {\sqrt{1 - \beta^2} \over \beta} \,.
\end{eqnarray}
We note that in this presentation the Melvin parameter $\beta$ appears exclusively in the scale $R_{\rm CTC}$; the limit $\beta \to 0$ returning to global AdS corresponds to $R_{\rm CTC} \to \infty$.    Unlike the Schr\"odinger case, where $\beta$ can be absorbed into a redefinition of the light-cone coordinates (\ref{LightCone}), here $\beta$ defines a radial scale which has physical meaning relative to the unit AdS radius.  It is natural to expect that this ratio will be a nontrivial parameter in a spacetime dual to a theory living on a sphere, which possesses an extra length scale corresponding to the sphere's radius.

We will call this spacetime the five-dimensional G\"odel-Schr\"odinger geometry, or ${\cal GS}_5$, though so far we have not justified the connection either to the Schr\"odinger solution ${\cal S}_5$ nor to G\"odel spacetimes.  In the remaining parts of this section we will do the former, while the latter will follow in the next section.

\subsection{Symmetries of the G\"odel-Schr\"odinger geometry}

As already described, five-dimensional AdS space has an $SO(4,2)$ symmetry group, where in the global coordinate system the natural subgroups $SO(4) \times SO(2)$ act in the usual way on the $S^3$ and the global time coordinate $\tau$.  The remaining 8 generators span a $({\bf 4}, {\bf 2})$ of $SO(4) \times SO(2)$; viewing $SO(4)$ as $SO(3) \times SO(3)$, they appear as a $({\bf 2}, {\bf 2}, {\bf 2})$ of $SO(3) \times SO(3) \times SO(2)$.

The null Melvin map breaks a number of these symmetries; however, nine of the fifteen isometries of global AdS in fact survive, the same number that is present for the Schr\"odinger space ${\cal S}_5$.  $SO(4) \simeq SO(3) \times SO(3)$ is broken, but since the new angular term in (\ref{MGAdS}) is proportional to the Hopf fiber $d\psi + \cos \theta d\phi$ in the fibration of $S^3$ over $S^2$, the $SO(3)$ acting on the $S^2$ is preserved, as is the $SO(2)$ of the Hopf coordinate $\psi$.  Thus $SO(4) \to SO(3) \times SO(2)$; the $SO(2)$ of global time translations is preserved as well.  This $SO(3) \times SO(2) \times SO(2)$ is generated by $\{ L_i, H, N\}$:
\begin{eqnarray}
\label{SymGens}
H &=&{1 \over 2} (\partial_\tau + 2 \partial_\psi) \equiv \partial_t \,, \quad \quad
N = 2 \partial_\psi - \partial_\tau \equiv  \partial_\xi \,, \quad \quad  L_3 = \partial_\phi \,, \\
L_1 &=&\sin \phi \partial_\theta - \cos \phi \csc \theta \partial_\psi  + \cos \phi \cot \theta \partial_\phi \,,  \quad \quad
L_2 =\cos \phi \partial_\theta + \sin \phi \csc \theta \partial_\psi  - \sin \phi \cot \theta \partial_\phi \,,
\nonumber
\end{eqnarray}
where we have defined the light cone coordinates
\begin{eqnarray}
t \equiv  \tau + y  \,, \quad \quad \xi \equiv
{1 \over 2} \left( y - \tau \right) \,,
\end{eqnarray}
with $y = \psi/2$.

The remaining four isometries are a subset of the eight ``off-diagonal" generators in global AdS, forming a $({\bf 2}, {\bf 2})$ of $SO(3) \times SO(2)_H$, while being invariant under $SO(2)_N$; as with ${\cal S}_5$, $N$ commutes with all generators.  They are most compactly expressed in the Boyer-Lindquist coordinates
\begin{eqnarray}
\label{BLCoords}
\theta_B = {\theta \over 2} \,, \quad \quad
\psi_B = {1 \over 2} (\psi + \phi) \,, \quad \quad
\phi_B = {1 \over 2} ( \phi - \psi) \,,
\end{eqnarray}
as
\begin{eqnarray}
K_1 = k_1 + k_4 \,, \quad K_2 = k_3 - k_2 \,, \quad K_3 = k_2 + k_3 \,, \quad K_4 = k_1 - k_4 \,,
\end{eqnarray}
where
\begin{eqnarray}
k_1 &=& {r \cos(\tau - \phi_B) \sin \theta_B \over \sqrt{1+r^2}} \partial_\tau + \sqrt{1 + r^2} \sin \theta_B \sin (\tau - \phi_B) \partial_r  \\ && + {\sqrt{1+r^2} \over r} \cos \theta_B \sin (\tau - \phi_B) \partial_{\theta_B} - {\sqrt{1 + r^2} \over r} \csc \theta_B \cos (\tau - \phi_B) \partial_{\phi_B} \,,\nonumber
\\
k_2 &=& {-r \sin(\tau - \phi_B) \sin \theta_B \over \sqrt{1+r^2}} \partial_\tau + \sqrt{1 + r^2} \sin \theta_B \cos (\tau - \phi_B) \partial_r  \\ && + {\sqrt{1+r^2} \over r} \cos \theta_B \cos (\tau - \phi_B) \partial_{\theta_B} + {\sqrt{1 + r^2} \over r} \csc \theta_B \sin (\tau - \phi_B) \partial_{\phi_B} \,,\nonumber \\
k_3 &=& {r \cos(\tau + \psi_B) \cos \theta_B \over \sqrt{1+r^2}} \partial_\tau + \sqrt{1 + r^2} \cos \theta_B \sin (\tau + \psi_B) \partial_r  \\ && - {\sqrt{1+r^2} \over r} \sin \theta_B \sin (\tau + \psi_B) \partial_{\theta_B} + {\sqrt{1 + r^2} \over r} \sec \theta_B \cos (\tau + \psi_B) \partial_{\psi_B} \,,\nonumber \\
k_4 &=& {r \sin(\tau + \psi_B) \cos \theta_B \over \sqrt{1+r^2}} \partial_\tau - \sqrt{1 + r^2} \cos \theta_B \cos (\tau + \psi_B) \partial_r  \\ && + {\sqrt{1+r^2} \over r} \sin \theta_B \cos (\tau + \psi_B) \partial_{\theta_B} + {\sqrt{1 + r^2} \over r} \sec \theta_B \sin (\tau + \psi_B) \partial_{\psi_B} \nonumber \,.
\end{eqnarray}
We note that all 9 generators are independent of $\beta$, and so go over smoothly to AdS generators in the $\beta \to 0$ limit.

The commutators not determined by the symmetry assignments given above are the mutual commutators of the $K$s, given by
\begin{eqnarray}
\label{KComm}
[K_1, K_2] = -4 L_3 \,, \quad & \quad [K_1, K_3] = - 6 H + N - 4 L_2 \,, \quad&\quad [K_1, K_4] = 4 L_1 \,, \\
{}[K_2, K_3] = 4 L_1 \,, \quad & \quad [K_2, K_4] = - 6 H + N + 4 L_2 \,, \quad&\quad [K_3, K_4] = - 4 L_3 \,. \nonumber
\end{eqnarray}
We see that $N$ appears as a central generator, as in the Schr\"odinger case, but $H$ and the $L_i$ appear on the right-hand-sides as well.

In Schr\"odinger holography, the space on which the dual field theory lives is the space obtained keeping both the radial coordinate and the coordinate associated to $N$ fixed.  Taking $r=$ constant, $\xi=$ constant we have a copy of $S^2 \times {\mathbb R}$ spanned by $\theta$, $\phi$ and $t$.  Meanwhile, the ${\cal GS}_5$ symmetry algebra based on  $SO(3) \times SO(2) \times SO(2)$  is quite similar to the Schr\"odinger algebra based on  $SO(2,1) \times SO(2) \times SO(2)$, but with differences appropriate to the dual theory living on the new space.
The Schr\"odinger $SO(2,1)$ (generated by $H$, $C$ and $D$) has been replaced by the spatial $SO(3)$ of the $S^2$; in the Schr\"odinger case the Hamiltonian is part of $SO(2,1)$ while $M_{12}$ forms an $SO(2)$, while in the G\"odel-Schr\"odinger case, it is the Hamiltonian generator that forms the $SO(2)$, describing the time direction of the dual.  Analogously to the case comparing Poincar\'e and global AdS, the generators that are ``spatial" relative to the boundary geometry are reshuffled amongst those corresponding to conformal enhancement.  It is thus natural to interpret the boundary CFT as living on $S^2 \times {\mathbb R}$ on which $SO(3) \times SO(2)_H$ naturally acts.

The other difference between the two symmetry algebras is the commutators of other four generators.  In both cases these four form a $({\bf 2}, {\bf 2})$ of $SO(2) \times SO(3)$ or $SO(2) \times SO(2,1)$; however in the ${\cal S}_5$ case, their mutual commutators include only the central element $N$, while in the ${\cal GS}_5$ case, the other generators are present on the right-hand side as well.

The large  isometry group is transitive, guaranteeing that the spacetime is homogeneous;  any point can be reached from any other by the application of symmetry transformations, implying that all points are equivalent.  One can see this by noting the $K$ generators may be used to translate in $r$, then at fixed $r$ one may move in $\theta$ using $L_1$ and $L_2$, before reaching any desired $\tau$, $\psi$, $\phi$ using $H$, $N$ and $L_3$.

\subsection{Schr\"odinger limit}

It is possible to take a coordinate limit of the G\"odel-Schr\"odinger metric to obtain the original Schr\"odinger solution; the same sort of limit works in obtaining the Poincar\'e patch representation of AdS from the global solution.  One imagines taking a point at large $r$ so the three-spheres are large compared to the AdS radius; recall we have set this radius to unity.  One must then look at small angle fluctuations, so the complete three-sphere is no longer visible.  Geodesics fall out of the Poincar\'e patch in finite proper time, so we consider small fluctuations of $\tau$ as well.  Finally, to preserve the Schr\"odinger character and not revert to ordinary anti-de Sitter space, we take the Melvin length scale $R_{\rm CTC}$ to be large such that $r/R_{\rm CTC}$ remains finite, implying $R_{\rm CTC}$ is much larger than the AdS scale, which also characterizes the radius of the sphere on which the dual NRCFT lives.

This can be implemented with the coordinate redefinition
\begin{eqnarray}
r = {r' \over \epsilon} \,, \quad \tau = \epsilon \tau'  \,, \quad \theta = {\pi \over 2} + 2 \epsilon x' \,,\quad
\psi = 2 \epsilon y' \,, \quad \phi = 2 \epsilon z' \,, \quad \beta = \epsilon \beta' \,,
\end{eqnarray}
and light-cone coordinates in  Schr\"odinger may be assembled in the usual way,
\begin{eqnarray}
t' \equiv \beta'(\tau' + y')\,, \quad \quad \xi' \equiv (y' - \tau')/2\beta' \,,
\end{eqnarray}
Substituting into the G\"odel-Schr\"odinger metric (\ref{MGAdS}) and taking the limit $\epsilon \to 0$, and finally dropping all primes in the end, returns precisely the ${\cal S}_5$ Schr\"odinger metric (\ref{PSchro}).  Since the G\"odel-Schr\"odinger spacetime is homogeneous, there is nothing special about the point we consider, and any point reduces locally to the spacetime ${\cal S}_5$ when $R_{\rm CTC} \gg 1$.

One naturally expects that the nine isometries of G\"odel-Schr\"odinger go over into the original Schr\"odinger isometries, and indeed this is the case.   Looking only at leading behavior in the 
small $\epsilon$ limit, only the isometries of mass dimension 1 appear at ${\cal O}(\epsilon^{-1})$:
\begin{eqnarray}
H &\to& {\beta' H' \over \epsilon}   \,, \quad
N \to {N' \over \beta' \epsilon} \,, \quad
L_2 \to {P'_x \over 2 \epsilon} \,, \quad
L_3 \to {P'_z \over 2 \epsilon} \,, \quad
L_1 \to - {\beta' H' \over 2 \epsilon} - {N' \over 4 \beta' \epsilon}\,,
\end{eqnarray}
with the $k$'s also giving various combinations of $H'$, $N'$, $P_x'$ and $P_z'$.
The remaining generators are present, but at higher order; those of mass dimension 0 appear at ${\cal O}(\epsilon^0)$, and the one of mass dimension $-1$ at ${\cal O}(\epsilon^1)$.  We can extract them all by taking appropriate linear combinations:
\begin{eqnarray}
H' &=& {\epsilon \over \beta'} H \,, \quad \quad N' = {\beta' \epsilon} N \,, \quad \quad
P_x' =  \lim_{\epsilon \to 0} 2 \epsilon\,  L_2 \,, \quad \quad P_z' = 2\,  \epsilon L_3 \,,  \\
M_{xz}' &=&  \lim_{\epsilon \to 0}  \left( L_1 + {1 \over 2} H + {1 \over 4} N \right) \,, \quad \quad
D' =  \lim_{\epsilon \to 0} 
{1 \over \sqrt{2}} (k_2 - k_4) \,,\\
K_x' &=& \lim_{\epsilon \to 0} \beta'  
\left( {1 \over \sqrt{2}} ( k_3 - k_1)
- 2 L_3 \right) \,, \quad \quad 
K_z' = \lim_{\epsilon \to 0} \beta'  
\left( {1 \over \sqrt{2}}(-k_2 - k_4) + 2 L_2 \right) \,, \\
C' &=& \lim_{\epsilon \to 0} {\beta' \over \epsilon}  
\left( {1 \over \sqrt{2}} (-k_1 - k_3)
- L_1 + {3 \over 2} H - {1 \over 4} N \right) \,,
\end{eqnarray}
Dropping the primes, these are precisely the generators (\ref{SchroIsoms}) of Schr\"odinger spacetime from section~\ref{MelvinSec}.  Note that the algebra is modified by the limit, since the ${\cal S}_5$ and ${\cal GS}_5$ symmetry groups are not the same; for example, the three-sphere rotations $L_2$ and $L_3$ do not commute in ${\cal GS}_5$, but in the limit taken, they become commuting translations $P_i$ in ${\cal S}_5$.

Thus we have justified that the G\"odel-Schr\"odinger spacetime has the appropriate symmetry algebra and limit to holographically encode the same NRCFT as ${\cal S}_5$, but on the space $S^2 \times {\mathbb R}$.

We note that the Schr\"odinger geometry ${\cal S}_5$ is known to be nonsupersymmetric \cite{Maldacena:2008wh}.  Given the limit relating them just described, this suggests that the G\"odel-Schr\"odinger spacetime may lack supersymmetry as well; however, this has not been explicitly verified.  There are also known to be generalizations of the Schr\"odinger geometry with unbroken supersymmetry \cite{Hartnoll:2008rs, Bobev:2009mw, Donos:2009xc, Ooguri:2009cv}.  Given the similarity between ${\cal GS}_5$ and several supersymmetric geometries described in the following sections, and the local stability associated to supersymmetric solutions, it would be useful to perform this analysis more explicitly.

\section{G\"odel spacetimes and holographic screens}
\label{ScreenSec}

So far we have argued for the similarities between this G\"odel-Schr\"odinger geometry and the Schr\"odinger space ${\cal S}_5$; we shall now justify the ``G\"odel" part of the name by demonstrating the existence of closed timelike curves in close analogy to known G\"odel-type spacetimes, and investigate the holographic screens of the geometry.

\subsection{Closed timelike curves}

We wish to demonstrate that the G\"odel-Schr\"odinger geometry possesses closed timelike curves (CTCs).  This is easy to do.  The $\phi$-direction is naturally periodic with periodicity $2 \pi$.  Moving in this direction only on the ${\cal GS}_5$ geometry, we have
\begin{eqnarray}
ds^2 \to G_{\phi\phi} d\phi^2  &=& 
{r^2 \over 4 } \left( 1 - {r^2 \over R_{\rm CTC}^2 } \cos^2 \theta \right) d \phi^2 \,.
\end{eqnarray}
This is manifestly spacelike at small $r$, and spacelike everywhere for the global AdS limit of $R_{\rm CTC} \to \infty$.  However, in the G\"odel-Schr\"odinger spacetime the direction becomes timelike at
\begin{eqnarray}
r^2 >  {R_{\rm CTC}^2 \over \cos^2 \theta} \,.
\end{eqnarray}
The angles giving the smallest value of the right-hand-side are $\theta = 0, \pi$.  (Note that the $d\phi$ direction is not degenerate within the three-sphere at these points; instead one of the Boyer-Lindquist coordinates  (\ref{BLCoords}) $\phi - \psi$ or $\phi + \psi$ becomes degenerate, with the other linear combination giving an $S^1$, which is the closed timelike curve we discover.)  Thus the transition radius to a zone with CTCs is 
\begin{eqnarray}
r > R_{\rm CTC} = {\sqrt{1- \beta^2} \over \beta} \,,
\end{eqnarray}
the radial scale parameter we defined previously (\ref{DefineRCTC}), justifying the subscript.
This radius obviously moves off to infinity as we take $\beta \to 0$.

Of course, the spacetime is homogeneous, so there are closed timelike curves going through every point.  What the argument above demonstrates is that these CTCs have a certain minimum size in the radial coordinate; speaking loosely, only a region of space of characteristic size $R_{\rm CTC}$ or larger will be able to sample a complete closed timelike curve.

The fact that the spacetime is homogeneous but possesses closed timelike curves is highly reminiscent of the original G\"odel spacetime, as well as its higher-dimensional descendants. All these geometries are naively pathological due to the closed timelike curves.  However, a number of ideas for rescuing or reinterpreting the various G\"odel solutions have been put forward, and we can apply these to our geometry as well.  In the remainder of this section, we will consider the calculation of a holographic preferred screen for an observer at the origin in G\"odel-Schr\"odinger spacetime, and show that it occurs in the region free from closed timelike curves, suggesting that the holographic description of the observer's dynamics is independent of the CTCs.  In the next section, we consider how a condensation of brane giant gravitons may modify the acausal region of the solution to remove the CTCs altogether.

\subsection{Comparison to the G\"odel and  suspersymmetric G\"odel geometries}

The original G\"odel solution \cite{Godel:1949ga} is a four-dimensional geometry solving Einstein's equations in the presence of pressureless dust and a negative cosmological constant.  The metric may be written
\begin{eqnarray}
ds^2 = {2 \over \omega^2} \left( - dt^2 + dr^2 - (\sinh^4 r - \sinh^2 r) d \phi^2 + 2 \sqrt{2} \sinh^2 r d \phi d t \right) + dz^2 \,,
\end{eqnarray}
with periodic coordinate $\phi$ \cite{Hawking:1973uf}.  The geometry is homogeneous, and orbits of the $\phi$ coordinate become closed timelike curves for $r > \log (1 + \sqrt{2})$. Although the G\"odel spacetime has a global time coordinate $t$, it has no global time function, as the constant-$t$ surfaces are not everywhere spacelike.

Certain properties of the G\"odel universe, notably the homogeneity and closed timelike curves, are also present in the spacetime we are considering.  An even closer cousin is the five-dimensional supersymmetric generalization of the G\"odel spacetime, discovered by Gauntlett, Gutkowski, Hull, Pakis and Reall \cite{Gauntlett:2002nw}.  This is a solution to five-dimensional minimal supergravity with metric
\begin{eqnarray}
ds^2 = - (d\tau + \beta' r^2 (d\psi + \cos \theta d \phi))^2 + dr^2 + r^2 d \Omega_3^2 \,, 
\end{eqnarray}
where $\beta'$ is the rotation parameter, along with a nontrivial massless gauge field.  Following \cite{Boyda:2002ba}, we shall refer to this spacetime as ${\cal G}_5$.  It may be lifted to a solution of eleven-dimensional supergravity, and is T-dual to a type IIB pp-wave spacetime.

This geometry shares a number of features with our G\"odel-Schr\"odinger spacetime, the most immediately obvious being the cross-term between the time coordinate $\tau$ and the Hopf fiber of the $S^3$.  They are both homogeneous, and also share the existence of closed timelike curves, which are present for ${\cal G}_5$ starting at a radius
\begin{eqnarray}
r   >  {1 \over 2 \beta'} \,.
\end{eqnarray}
Furthermore, the symmetry groups are virtually identical.  Both possess a nine-dimensional group of isometries consisting of $SO(3) \times SO(2) \times SO(2)$ along with four more generators transforming in the ${\bf (2,2,1)}$.  The only difference is that for ${\cal G}_5$ these four generators commute except for closing onto the central extension, as was the case for ${\cal S}_5$, in contrast with the ${\cal SG}_5$ case, where the $SO(3) \times SO(2)$ generators appear on the right-hand sides as well, as shown in equation~(\ref{KComm}).  

Other differences are that ${\cal G}_5$ is supported by a massless gauge field and reverts to Minkowski space as the rotation parameter is set to zero, while ${\cal SG}_5$ features a massive gauge field and a cosmological constant, and becomes global AdS in the non-rotating limit.  Thus in a sense, our geometry is an anti-de Sitter version of the G\"odel spacetime.  There is also the question of whether ${\cal GS}_5$ is indeed nonsupersymmetric, and if so, whether there exist other solutions with properties analogous to ${\cal GS}_5$ (in particular, closed timelike curves and asymptotic AdS space) that possess supersymmetry as well.

\subsection{Holographic screens in the supersymmetric G\"odel universe}

The covariant holographic principle of Bousso \cite{Bousso:1999xy,Bousso:1999cb} states that the size of a codimension-two area $A$ bounds the entropy (or degrees of freedom) passing through a ``light sheet" generated by the contracting null rays orthogonal to $A$.   In general two of the four possible null congruences orthogonal to $A$ (two future-going, two past-going) will be shrinking; in weak gravity this is the ``inward"-moving congruence in each the past and the future, but this can change in the presence of strong gravity.  A light-sheet terminates when the expansion ceases being negative; when the null-energy condition is satisfied, this can only happen at a caustic where the curves pass through each other.

For a given observer, one may take the families of null rays emanating from  any point on the observer's worldline and follow them until they cease expansion; the area $A$ that is the locus of zero expansion is called a ``preferred screen", with the family of null rays forming a light sheet for the screen.  The union of all the observer's preferred screens forms a hypersurface called the preferred screen-hypersurface; generally this hypersurface is simply referred to as a preferred screen itself, and we will follow this convention.  The preferred screen holographically encodes a region consisting of all the light sheets associated to the null congruences, including the observer's worldline; the region so encoded is called the holographic domain of the screen.

For anti-de Sitter space and similar spacetimes, the preferred screen for all observers is the boundary, and the holographic domain of the boundary is all of spacetime; thus there is a universal, observer-independent holographic boundary theory describing the entire geometry.  This is not necessarily the case for a general spacetime.  Notably, for de Sitter space, each individual observer has a distinct preferred screen coinciding with that observer's cosmological horizon, whose holographic domain does not span the entire space but instead is of finite size, suggesting each observer only has access to a finite number of degrees of freedom.  Another example in an AdS-type space is the Karch-Randall geometry of an AdS$_4$ brane inside AdS$_5$ \cite{Karch:2000ct}, where part of the space is in the holographic domain of the AdS$_4$ brane, and the rest is in the domain of the remainder of the AdS$_5$ boundary \cite{Bousso:2001cf}.  One expects that for a general geometry, holography will be observer-dependent in some way.

As discussed in \cite{Boyda:2002ba}, holography in G\"odel-type universes is 
an example of such observer-dependence.
  For the supersymmetric G\"odel universe, congruences of null rays expanding from the observer at $r=0$ do not expand to infinity but instead reach precisely the radius $r = 1/(2 \beta')$ where closed timelike curves appear, called there the velocity-of-light surface, before contracting again.  The surface of zero expansion, that is the preferred screen for the $r=0$ observer, was found to be at
\begin{eqnarray}
\label{GScreen}
r = {\sqrt{3} \over 4 \beta'} = {\sqrt{3} \over 2} R_{\rm CTC} \,.
\end{eqnarray}
By homogeneity, this analysis holds for any point in the space.
Thus no single observer's preferred screen contains a complete closed timelike curve within its holographic domain, suggesting that the dynamics experienced by this observer may be consistently described by a theory living on the screen that is ignorant of the existence of the CTCs; for further discussion, see \cite{Boyda:2002ba}.

\begin{figure}
  \centerline{\includegraphics{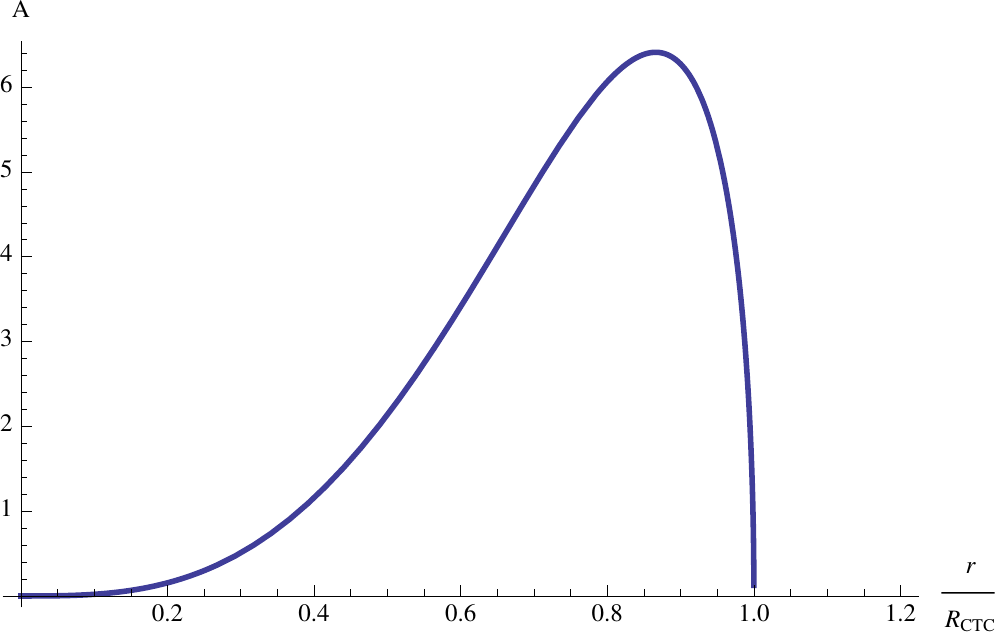}}
    \caption{The area of the constant-$\tau$, constant-$r$ slice of the G\"odel-Schr\"odinger spacetime.  It peaks at the screen location $r/R_{\rm CTC} = \sqrt{3}/2$ before going to zero at $r= R_{\rm CTC}$.}\label{AreaFig}
 \end{figure}

We will now demonstrate that a completely analogous story obtains for the G\"odel-Schr\"odinger spacetime.  Consider an observer at $r=0$; again, homogeneity will imply the same obtains for any other point. 
The simplest way to find the screen is to note that 
given the large symmetry group of the ${\cal GS}_5$ spacetime, in particular the preserved $SO(3) \times SO(2)$ acting on the squashed three-sphere, we may infer that the preferred screen will be an $r=$ constant hypersurface.  (We will demonstrate a more rigorous method momentarily.) Taking
the metric on a slice $\tau =$ constant, $r =$ constant,
\begin{eqnarray}
ds^2 = {r^2 \over 4} \left( {1 - \beta^2 - \beta^2 r^2 \over 1 - \beta^2} (d \psi + \cos \theta d\phi)^2 + d\theta^2 + \sin^2 \theta d\phi^2 \right) \,,
\end{eqnarray}
which is just the metric of the squashed three-sphere, we may calculate the area as a function of $r$, which is
\begin{eqnarray}
A = \int \sqrt{g_3} = 2 \pi^2 r^3 \sqrt{1 - {r^2 \over R_{\rm CTC}^2} }  \,,
\end{eqnarray}
which as $R_{\rm CTC} \to \infty$ is just the volume of the ordinary three-sphere.  Starting from the observer at $r=0$, the area increases until one reaches the maximal area, which is the location of the screen:
\begin{eqnarray}
\label{Screen}
r_{\rm screen} = {\sqrt{3} \over 2} R_{\rm CTC} \,.
\end{eqnarray}
For larger $r$ the area decreases until it goes to zero at $R_{\rm CTC}$, heralding the appearance of closed timelike curves as $\tau$ fails to be a global time function; see figure~\ref{AreaFig}.
We note that the relationship between the screen location and the CTC radius is numerically identical to  that for the SUSY G\"odel spacetime (\ref{GScreen}), further demonstrating the similarity between these two geometries.

One expects that a similar argument would work for more general spaces with closed timelike curves, and sufficient symmetry.  Whenever the area of the would-be screen contains the direction that develops the closed timelike curve, the onset of CTCs will send the area to zero.  Given the area is also zero at the origin of the radial coordinate and growing as one moves away, there must be a maximal area somewhere before the CTC radius, which is the location of the screen.

To check that this is the correct result for the location of the preferred screen, and to better understand the geometry, we may explicitly find the null congruences heading away from the $r=0$ observer and calculate where their expansion vanishes.
Consider null geodesics passing through the origin; let a given geodesic be parameterized by $U^\mu \equiv dx^\mu / d\lambda$ with $\lambda$ the affine parameter.  Three constants of the motion  are immediately obtainable from the fact that the coordinates $\tau$, $\psi$ and $\phi$ are absent from the metric:
\begin{eqnarray}
E \equiv - U_\tau \,, \quad \quad L_\psi \equiv U_\psi \,, \quad \quad L_\phi \equiv U_\phi \,,
\end{eqnarray}
in terms of which we can solve for $U^\tau$, $U^\psi$ and $U^\phi$:
\begin{eqnarray}
{d \tau \over d \lambda } &=& {E \over 1 + r^2} - \beta^2 (E + 2 L_\psi) \,, \\
{d \psi \over d \lambda } &=& {4 \over r^2\sin^2 \theta} (L_\psi  - L_\phi \cos \theta) + 2 \beta^2 (E + 2 L_\psi) \,, \\
{d \phi \over d \lambda } &=&  {4 \over r^2\sin^2 \theta} (L_\phi - L_\psi \cos \theta)\,.
\end{eqnarray}
The norm of the tangent vector to the geodesic is then
\begin{eqnarray}
g_{\mu\nu} U^\mu U^\nu &=& { (L_\psi + L_\phi)^2 \over r^2 \cos^2 (\theta/2)} + {(L_\psi - L_\phi)^2 \over r^2 \sin^2 (\theta/2)} \\
&+&  {1 \over 1+ r^2}\Big(  (U^r)^2 + \beta^2 (E + 2 L_\psi)^2 (1 + r^2) - E^2 \Big) + {r^2 (U^\theta)^2 \over 4} \,. \nonumber
\end{eqnarray}
This must vanish for a null geodesic; given the first two terms, this is only possible at $r=0$ if $L_\psi = L_\phi = 0$.  Thus as one would expect for a geodesic passing through the origin, both angular momenta are zero.  For a null geodesic, we may also rescale the affine parameter to impose $E = 1$.

Since the total $SO(3)$ angular momentum is zero, the rotational symmetry generators $L_1$ and $L_2$ (\ref{SymGens}) must also have zero inner product with the geodesic  tangent vector.  Requiring this for arbitrary $r$ and $\theta$ forces $U^\theta = 0$.  The only remaining component is then $U^r$, and solving for it gives us
\begin{eqnarray}
U^r = \sqrt{R_{\rm CTC}^2 - r^2 \over 1 + R_{\rm CTC}^2} \,, \quad \quad
U^\tau = {1 \over1 + r^2} - \beta^2 \,, \quad \quad U^\psi = 2 \beta^2 \,, \quad \quad
U^\theta = U^\phi =0 \,.
\end{eqnarray}
It is already apparent  a solution will only exist for $r  \leq R_{\rm CTC}$.  One can solve the resulting equations to find the solution for null geodesics passing through the origin,
\begin{eqnarray}
{r \over R_{\rm CTC}} = \sqrt{1 \over 1 + \cot^2 \beta \lambda} \,,\quad \quad  \psi = \psi_0 + 2 \beta^2 \lambda \,, \quad \quad \tau = \tau_0 + \tan^{-1} \left( \tan \beta \lambda \over \beta \right) - \beta^2 \lambda \,,
\end{eqnarray}
with $\theta$ and $\phi$ constant, where $\psi_0$ and $\tau_0$ are arbitrary initial values and  we took $r = 0$ at $\lambda =0$.  Since the spacetime is homogeneous, geodesics passing through any point behave the same way.

\begin{figure}
  \centerline{\includegraphics{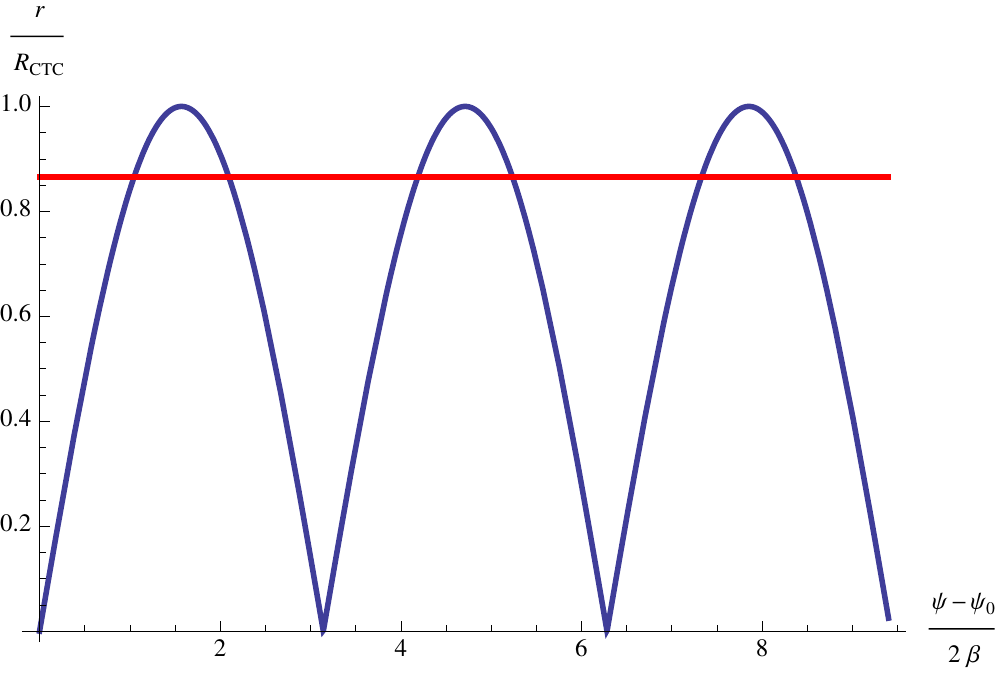}}
    \caption{The radial motion of the null geodesics passing through the origin plotted against their rotation in the $\psi$ direction, which is directly proportional to the affine parameter.  The radius of the preferred screen is the red horizontal line.}\label{GeodesicsFig}
 \end{figure}

Thus as expected for a G\"odel universe, null geodesics spiral outward, advancing in both $r$ and $\psi$ as time progresses, while staying fixed on the two-sphere parameterized by $\theta$ and $\phi$, until they reach a maximum radius of $r = R_{\rm CTC}$, after which they begin contracting again.  They contract in  fashion symmetric to the expansion until they reach $r=0$, where the process repeats.  The radial oscillation as a function of the rotation in $\psi$ (or equally well the affine parameter) is displayed in figure~\ref{GeodesicsFig}.

We may finally calculate the expansion $\theta$ of the null geodesic congruence, whose vanishing defines the preferred screen:
\begin{eqnarray}
\theta \equiv \nabla_\mu U^\mu = {3 - (3 + 4 r^2) \beta^2 \over r  \sqrt{1 - \beta^2 (1 + r^2)}}
 = {3 R_{\rm CTC}^2 - 4 r^2 \over 2 \sqrt{(1 + R_{\rm CTC}^2)(R_{\rm CTC}^2 - r^2)}} \,,
\end{eqnarray}
which goes to zero at
\begin{eqnarray}
r = {\sqrt{3} \over 2} R_{\rm CTC} \,,
\end{eqnarray}
precisely the location of maximal area;  this  indeed agrees with our previous result for the location of the preferred screen.

It is interesting to note how G\"odel-Schr\"odinger inherits properties both of anti-de Sitter space as well as the G\"odel-type geometries.  In AdS, massive geodesics through the origin move out to an energy-dependent maximum radius before re-collapsing, while the lightlike geodesics make it all the way to the boundary; in a sense AdS's primary characteristic is that it traps massive particles but allows massless ones to reach infinity in finite time.  In G\"odel spacetimes, in contrast, the null geodesics are trapped as well, and re-collapse to the origin in finite time, while rotating.  This is another way of describing the observation that anti-de Sitter space is holographically represented by a screen at the boundary, while G\"odel-type universes have a holographic screen (for the observer at the origin) at a finite radius.

In sum, we have demonstrated the close kinship between the G\"odel-Schr\"odinger spacetime and the supersymmetric G\"odel universe ${\cal G}_5$; the original G\"odel universe has a similar set of preferred screens \cite{Boyda:2002ba}, as will other, similar universes.  The existence of the finite-sized preferred screen implies that any individual observer's dynamics may be calculable without having to include the CTCs, which are too large to fit inside any single holographic domain.  This offers a kind of chronology protection.

It may not be fully satisfactory, however, since it still leaves us with a global geometry in which the CTCs are present.  One may wonder whether string theory directly addresses their existence, or only skirts it in the holographic description.  In the next section, we will argue that a condensation of branes should change the geometry in the region with closed timelike curves, presumably removing them, without affecting the holographic domain of the preferred screen.

\section{Giant gravitons in G\"odel-Schr\"odinger and the repulson}
\label{BraneSec}

Hawking \cite{Hawking:1991nk} proposed the existence of a Chronology Protection Conjecture, that  closed timelike curves are prevented from developing in geometries that begin without them; see also for example \cite{Deser:1991ye}.  It is another question, however, to ask whether string theory or any other theory of quantum gravity might ``resolve" closed timelike curves in spacetimes that classically possess them at all times.  It was suggested by Dyson \cite{Dyson:2003zn} that in certain cases where the classical geometry has CTCs, the dynamics of branes may generate an enhan\c{c}on-type mechanism that changes the geometry to remove the acausal behavior; in her example, the closed timelike curves lived within a compact region, with chronologically safe geometry at infinity.  A case closer to ours, where the CTCs appear outside a certain radius, was provided by  Drukker, Fiol and Sim\'on (DFS) \cite{Drukker:2003sc}, involving a condensation of supertubes in a four-dimensional G\"odel-type geometry.  One is naturally led to ask whether Dyson's enhan\c{c}on could operate in the G\"odel-Schr\"odinger system as well, and as we shall show, the answer is that it does.

\subsection{Four-dimensional G\"odel spacetime and supertubes}
Beyond the original supersymmetric G\"odel universe ${\cal G}_5$, a number of other G\"odel-type geometries that preserve some supersymmetry and appear as solutions to  supergravity equations of motion have been found.\footnote{For work applying the ideas used in this section to ${\cal G}_5$, see \cite{Gimon:2004if}.}  A geometry solving the equations of type IIA supergravity was  described by Harmark and Takayanagi \cite{Harmark:2003ud} with the form ${\cal G}_4 \times {\mathbb R}^6$, where the metric and $p$-form fields are trivial on the ${\mathbb R}^6$, and ${\cal G}_4$ has metric
\begin{eqnarray}
ds^2 = - (dt +  c r^2 d \phi)^2 + dr^2 + r^2 d\phi^2 + dy^2 \,,
\end{eqnarray}
and $p$-form field strengths
\begin{eqnarray}
H_3 = - 2 c r dr \wedge d\phi \wedge dy \,, \quad \quad
F_2 = -2 c r dr \wedge d\phi \,, \quad \quad
F_4 = 2 c r dt \wedge dr \wedge d\phi \wedge dy \,.
\end{eqnarray}
Again we have a G\"odel-type spacetime, with closed timelike curves manifesting themselves for $ r > 1/c$.  An analysis of the area of constant-$t$, constant $r$-slices implies a preferred screen for the observer at the origin at 
\begin{eqnarray}
r_{\cal S} = {1 \over \sqrt{2}c} \,,
\end{eqnarray}
again less than the critical radius for the appearance of CTCs.

As pointed out in  \cite{Dyson:2003zn, Drukker:2003sc}, a natural string theory probe of a geometry with closed timelike curves is  a brane, since the brane's finite extent is capable of wrapping the entire acausal curve.  DFS found that a suitable brane probe of the ${\cal G}_4$ geometry is a supertube \cite{Mateos:2001qs,Emparan:2001ux}.  A supertube is an object with D0-brane and fundamental string charge, as well as D2-brane dipole moment; it may be realized as a D2-brane with both electric $E$ and magnetic fields $B$ on its world volume, which act as sources for the F-string charge $q_s$ and D0-brane charge $q_0$ respectively.  For a tube filling $t$, $y$ and $\phi$ we have the worldvolume field strength 
\begin{eqnarray}
{\cal F} = E \, dt \wedge dy +(B - cr^2) \, dy \wedge d\phi \,,
\end{eqnarray}
and the associated charges,
\begin{eqnarray}
q_s = \int d \phi \, \Pi \,, \quad \quad q_0 = \int d \phi \, B \,,
\end{eqnarray}
with $\Pi$ the momentum conjugate to $E$.  For given conserved charges, the crossed worldvolume fields generate an angular momentum,
\begin{eqnarray}
J = q_s q_0 \,,
\end{eqnarray}
which stabilizes the tubular brane against collapse at a finite $r = R_J$\,,
\begin{eqnarray}
R_J \equiv \sqrt{|J|} \,.
\end{eqnarray}
The solution to the equations of motion has $|E|=1$ for any $R_J$.  Thus one can have a supertube at any radius by tuning the conserved charges appropriately.

Drukker, Fiol and Sim\'on considered a supertube embedded in the G\"odel universe ${\cal G}_4$ and found that the above analysis, originally performed in flat space, obtains equally well with the $-cr^2$ term added to the magnetic field.  The supertube is a source for the same fields that are turned on in the background, and preserves some supersymmetry, making it a natural probe to consider.
  However, considering radial fluctuations of the supertube around its minimum radius in the ${\cal G}_4$ geometry, they found kinetic terms
\begin{eqnarray}
{\cal L} = {\cal L}_0 + {1 \over 2} M \dot{r}^2 + \ldots \,,
\end{eqnarray}
with  coefficient
\begin{eqnarray}
M = {(B - c r^2)^2 + r^2 - c^2 r^4 \over B} \,,
\end{eqnarray}
which is positive-definite for $r < 1/c$, but passes through zero precisely at the CTC radius and becomes negative at larger $r$.

DFS argued that such a negative kinetic term is a signal that the moduli space metric no longer contains the proper degrees of freedom, analogous to the repulson singularity \cite{Behrndt:1995tr,Kallosh:1995yz,Cvetic:1995mx}.  Continuing the analogy, the repulson should be removed by the condensation of branes with negative kinetic terms, in this case the supertubes, and replaced by a new ``enhan\c{c}on" geometry \cite{Johnson:1999qt}.  DFS  suggested that the spacetime sourced by a collection of supertubes was the proper new geometry outside the CTC radius, and thus that after the condensation of branes the final geometry would be ${\cal G}_4$ for $r < 1/c$, and the supertube-backreacted geometry for $r > 1/c$.  The resulting spacetime is free of closed timelike curves.

It is then natural to ask whether an analogous ``repulson" instability exists in the G\"odel-Schr\"odinger spacetime.  We will show that this is indeed the case.  
The brane playing the role of the supertube will be a cousin, the dual giant graviton; appropriately enough for the ``rotating" G\"odel geometries, this probe also possesses angular momentum.  Before studying the giant graviton in ${\cal GS}_5$, we recall its description in global AdS.

\subsection{Giant gravitons in global AdS}

The original giant graviton solution in $AdS_5 \times S^5$ consists of a D3-brane propagating in time and wrapped on an $S^3 \subset S^5$ while moving at constant speed on an equator of $S^5$ transverse to the $S^3$ \cite{McGreevy:2000cw}.  Shortly thereafter, the ``dual" giant graviton solution was formulated, which wraps the $S^3$ in global $AdS_5$ instead, while propagating in global time and moving on an equator of $S^5$ \cite{Grisaru:2000zn,Hashimoto:2000zp}.  This solution may be seen as follows.

The action for a D3-brane, including only terms relevant in this paper, consists of the Born-Infeld and Chern-Simons terms,
\begin{eqnarray}
S = - \mu_3 \int d^4x \, e^{-\Phi} \sqrt{- \det( P[g - B])} + \mu_3 \int P[C_4] \,,
\end{eqnarray}
where $\mu_3$ is the brane charge and tension, and the symbol $P$ reminds us to pull back the spacetime fields to the brane worldvolume.  For the global AdS solution, we have the metric 
\begin{eqnarray}
ds^2 = ds^2_{\rm Global}  + ds^2_{S^5} \,,
\end{eqnarray}
with $ds^2_{\rm Global}$ given in (\ref{GlobalAdS}) and $ds^2_{S^5} = (d \chi + {\cal A})^2 + ds^2({\mathbb P}^2)$, $B=0$, a constant dilaton and 
\begin{eqnarray}
C_4 = {r^4 \over g_s} d\tau \wedge {\rm vol}_{S^3} + \ldots \,,
\end{eqnarray}
where the ellipsis indicates parts polarized entirely on the $S^5$ which are not relevant for us.  We consider a brane moving along the $S^5$ Hopf fiber with constant velocity,
\begin{eqnarray}
\omega \equiv {d \chi \over d \tau} \,.
\end{eqnarray}
Defining the Lagrangian ${\cal L}$ by
\begin{eqnarray}
S = {\mu_3\over g_s} \int d^4x \sqrt{-g_{S^3}} {\cal L} \,, 
\end{eqnarray}
and using that the pullback of the metric includes the term $P[g]_{\tau\tau} = g_{\tau \tau} + \omega^2 g_{\chi\chi}$, 
we find
\begin{eqnarray}
{\cal L} = r^4 - r^3 \sqrt{1 + r^2- \omega^2}\,.
\end{eqnarray}
Note that for the Poincar\'e case, $(1 + r^2)$ is replaced by $r^2$ and the potential vanishes at $\omega = 0$, leading to a moduli space for any $r$.  In the global case, one can find the conserved angular momentum,
\begin{eqnarray}
J \equiv {\partial {\cal L}\over \partial \omega}  = {\omega r^3 \over \sqrt{1 + r^2 - \omega^2}} \,,
\end{eqnarray}
and the Hamiltonian is
\begin{eqnarray}
{\cal H} \equiv J \omega - {\cal L} = \sqrt{(1 + r^2)(J^2 + r^6)} - r^4 \,.
\end{eqnarray}
This energy functional is minimized for given $J$ at either $r = 0$ (the trivial solution) or
\begin{eqnarray}
r = r_{\rm GG} \equiv \sqrt{|J|} \,. 
\end{eqnarray}
By choosing the appropriate angular momentum, a giant graviton solution can be set up at any $r$.  We note the close analogy to the supertube, where an extended brane is also stabilized at finite radius by angular momentum; the relationship $r^2 = |J|$  is identical in both cases.  Moreover, for a giant graviton solution one has $|\omega| = 1$, analogous to $|E|=1$ for the supertube.

\subsection{Giant gravitons in G\"odel-Schr\"odinger}

One can imagine taking a global AdS spacetime with a giant graviton in it, and performing the null Melvin twist.  The twist consists of coordinate redefinitions and two T-dualities.  Since the  Melvin map acting on the spacetime leaves $F_5$ unchanged and does not add any other RR fields, one might speculate that the D3-brane giant graviton (which is a source for $F_5$) might exist essentially unchanged in the G\"odel-Schr\"odinger spacetime.  We will find that this is true --- with the important caveat that giant gravitons localized in the region with CTCs have negative kinetic terms, and hence manifest a repulson-type sickness.

Rather than attempting to act the null Melvin twist on the D3-brane, we will instead consider directly D3-brane solutions in the G\"odel-Schr\"odinger spacetime.
The  ten-dimensional metric is now
\begin{eqnarray}
ds^2 = ds^2_{{\cal GS}_5} + {1 \over 1 - \beta^2} ds^2_{S^5} \,,
\end{eqnarray}
with $ds^2_{{\cal GS}_5}$ the 5D G\"odel-Schr\"odinger metric (\ref{MGAdS}).  The dilaton is
\begin{eqnarray}
\label{GSDilaton}
e^{\Phi} = { g_s \over \sqrt{1 - \beta^2}} \,,
\end{eqnarray}
and there is now a $B$-field
\begin{eqnarray}
B_2 =  - {\beta \over 1 - \beta^2} \left( (1 + r^2) d\tau + {r^2 \over 2} (d\psi + \cos \theta d \phi) \right) \wedge (d \chi + {\cal A}) \,.
\end{eqnarray}
We again assume a D3-brane wrapped on the (now squashed) $S^3$, propagating in $\tau$ and moving on $\chi$ with constant velocity $\omega$.  The pullback of the metric again includes a $g_{\chi\chi}$ term,
\begin{eqnarray}
P[g]_{\tau\tau} = g_{\tau \tau} + \omega^2 g_{\chi\chi} = g_{\tau\tau} + {\omega^2 \over 1 - \beta^2} \,.
\end{eqnarray}
The $B$-field has one index polarized in the D3-brane directions and one along the $\chi$ direction, resulting in a pullback that is nontrivial,
\begin{eqnarray}
P[B] = {\beta \omega \over 1- \beta^2} {r^2 \over 2} d\tau \wedge (d \psi + \cos \theta d \phi) \,.
\end{eqnarray}
Putting this together, we find the simple result,
\begin{eqnarray}
\sqrt{- \det( P[g - B])}  = {1\over  \sqrt{1 - \beta^2}} \left(r^3 \sqrt{1 + r^2 - \omega^2} \right) \,,
\end{eqnarray}
which when combined with the dilaton (\ref{GSDilaton}) leads to the conclusion that the Lagrangian for the giant graviton in the G\"odel-Schr\"odinger geometry is identical to the AdS case,
\begin{eqnarray}
{\cal L}_0 = r^4 - r^3 \sqrt{1 + r^2- \omega^2}\,.
\end{eqnarray}
Thus we can again have a solution at any $r_{\rm GG} = \sqrt{J}$ by choosing a suitable angular momentum.  One can trace this initially surprising simplification to the relationship (\ref{MetricAM}) between $B_2 \sim A_M$ and $\Delta ds^2 \sim A_M^2$.

Let us now turn to the radial fluctuations around such a giant graviton solution.  Adding a velocity $\dot{r} \equiv dr/d\tau$, we now obtain
\begin{eqnarray}
P[g]_{\tau\tau} = g_{\tau \tau} + \dot{r}^2 g_{rr} + \omega^2 g_{\chi\chi} \,,
\end{eqnarray}
we find the Lagrangian
\begin{eqnarray}
{\cal L} = r^4 - r^3 \sqrt{ 1 + r^2 - \omega^2 - { (1 - (1 + r^2)\beta^2) \dot{r}^2 \over 1 + r^2}} \,,
\end{eqnarray}
which expanding in powers of $\dot{r}$ becomes
\begin{eqnarray}
{\cal L} = {\cal L}_0 + {1 \over 2} M \dot{r}^2 + \ldots \,,
\end{eqnarray}
where we have dropped terms higher order in $\dot{r}^2$, and
\begin{eqnarray}
M = {r^3( 1 - (1 + r^2) \beta^2) \over (1 + r^2) \sqrt{1 + r^2 - \omega^2}}
= { \sqrt{J^2 + r^6}( 1 - (1 + r^2) \beta^2) \over (1 + r^2)^{3/2} }
 \,.
\end{eqnarray}
This is positive-definite for $\beta = 0$ (the global AdS case), and for nonzero $\beta$ it remains positive for small $r$.  However, the kinetic term coefficient becomes zero and then negative at $(1 + r^2) \beta^2 \geq 1$, or
\begin{eqnarray}
r \geq {\sqrt{1 - \beta^2} \over \beta} \equiv R_{\rm CTC} \,.
\end{eqnarray}
Thus precisely in the region where closed timelike curves are present, the branes develop a repulson-type instability, suggesting the solution given is not the true vacuum in that region.  Thus we expect the dual giant gravitons to condense and generate a new spacetime, presumably free of the chronological issues related to the presence of closed timelike curves.

\section{Conclusions}
\label{ConclusionSec}

We have used the null Melvin twist on a rotational isometry of anti-de Sitter space to generate a geometry with properties of both the G\"odel and Schr\"odinger spacetimes.  This G\"odel-Schr\"odinger spacetime is homogeneous with a large isometry group, appropriate to describe a non-relativistic conformal field theory on $S^2 \times {\mathbb R}$ and reducing to the Schr\"odinger geometry in a suitable limit.  This geometry possesses closed timelike curves, but we have argued that it may be useful nonetheless.  The preferred screen associated to an observer at the origin encloses a holographic domain small enough that no complete closed timelike curve is contained within it, suggesting the observer-dependent holographic dual description of this observer's experience is free of acausal pathology.  Meanwhile, a repulson-type instability in the dynamics of probe dual giant gravitons manifests itself precisely in the region with closed timelike curves, implying that the true geometry will be modified by brane condensation in this region, while leaving the holographic domain of the observer's screen intact.  This is further evidence that string theory engages in a kind of chronology protection, as described by Dyson \cite{Dyson:2003zn}, removing closed timelike curves by means of an enhan\c{c}on-type mechanism.

A number of questions naturally present themselves.  The most immediate is, what is the spacetime after giant graviton brane condensation, the G\"odel-Schr\"odinger version of the enhan\c{c}on? For the case of ${\cal G}_4$, evidence was given that the condensation of supertubes results in a domain wall solution, with ${\cal G}_4$ inside the wall, and the geometry sourced by  the supertube backreaction outside \cite{Drukker:2003sc}.  It is possible that a similar result occurs here, although we note a distinction between the cases:  the supertube is a source for all the relevant spacetime $p$-form gauge fields due to the electric and magnetic fields turned on, while our giant graviton has no worldvolume fields, and thus  does not seem to be a source for a $B_2$ field as found in the ${\cal GS}_5$ background.

A related question involves the action of the null Melvin twist, which is made out of coordinate shifts and T-dualities.  One would expect both of these operations to map proper solutions to proper solutions, and indeed, as a supergravity solution the G\"odel-Schr\"odinger spacetime is consistent.  It contains no strong curvatures or singularities and may exist at arbitrarily weak coupling, and thus does not appear to immediately suggest string theory corrections need to be added.  Indeed, the instability is nonlocal, involving the complete closed timelike curves and the extended branes, and thus the corrections go beyond string perturbation theory to the nonperturbative arena.
 Understanding better how the pathological properties of the G\"odel-Schr\"odinger solution arise from the action of T-dualities on anti-de Sitter space would shed light on the nature of how string theory relates to and resolves non-local pathologies, and to the nonperturbative nature of string theory in general.

It is also interesting to ask how the brane condensation relates to the homogeneity of the G\"odel-Schr\"odinger spacetime.  Since every point is equivalent, observers through every point perceive their own preferred screen beyond which closed timelike curves manifest.  The condensation of branes leading to a new enhan\c{c}on geometry appears to break this homogeneity.  Does the brane condensation indicate an actual spontaneous symmetry breaking, with a single observer becoming preferred, or does each observer continue to perceive their own slice of ${\cal GS}_5$, each of which being in some sense gauge equivalent to the others, in a manner analogous to horizon complementarity?  More understanding of these questions would be very useful in understanding how string theory implements holography and removes pathological features; for related thoughts, see \cite{Drukker:2003mg}.

The question of the relationship between supersymmetry and the G\"odel-Schr\"odinger space-time is also of interest.  The Schr\"odinger geometry ${\cal S}_5$ is nonsupersymmetric, but several of the G\"odel geometries we have described possess supersymmetry, as do generalizations of the Schr\"odinger spacetime \cite{Hartnoll:2008rs, Bobev:2009mw, Donos:2009xc, Ooguri:2009cv}.
The ability to reach ${\cal S}_5$ as a coordinate limit of  G\"odel-Schr\"odinger suggests the latter is nonsupersymmetric, but this should be verified explicitly.  If it is indeed nonsupersymmetric, one is naturally led to wonder whether the G\"odel-Schr\"odinger geometry, like the original G\"odel spacetime, admits generalizations (possessing closed timelike curves and anti-de Sitter asymptotics, for example) that are supersymmetric.

Certain G\"odel universes are known to be T-dual to pp-wave geometries \cite{Boyda:2002ba, Harmark:2003ud}.  A natural additional question is then whether the G\"odel-Schr\"odinger geometry has a dual description as a pp-wave or any other spacetime.  A class of pp-wave spacetimes can be obtained as limits of anti-de Sitter space and its cousins; since the null Melvin twist generates G\"odel-Schr\"odinger from AdS using coordinate limits and T-dualities, such a relationship would not be surprising.

Finally, it is interesting to speculate on the application of the closed timelike curves to Schr\"odinger holography.  The null Melvin twist acting on rotating black holes in AdS will generate G\"odel-like spacetimes dual to spinning NRCFTs that possess CTCs, and string theory presumably resolves them in a fashion analogous  to that described here.  One is led to wonder what, if any, meaning the closed timelike curves and their nonperturbative removal have in the non-gravitational dual.  These issues are of considerable interest for further exploration.

\section*{Acknowledgments}

We would like to thank A.~Adams, S.~de Alwis, S.~Kachru, I.~Klebanov, and C.~Rosen
for helpful discussions.
The work of C.M.B.\ and O.D.\ was supported by the Department of Energy under Grant
No.~DE-FG02-91-ER-40672.

\end{document}